\documentclass[reprint,superscriptaddress,amsmath,amssymb,aps,prl,floatfix]{revtex4-2}

\usepackage{graphicx}
\usepackage{dcolumn}
\usepackage{bm}
\usepackage[output-decimal-marker={.},exponent-product=\cdot]{siunitx}
\usepackage{hyperref}
\usepackage[mathlines]{lineno}
\usepackage{makecell}
\usepackage{color}
\usepackage[english]{babel}
\usepackage[T1]{fontenc}
\usepackage{multirow}
\usepackage{float}

\begin{document}


\title{Prominent bump in the two-neutron separation energies of neutron-rich lanthanum isotopes revealed by high-precision mass spectrometry}

\author{A.~Jaries}
\email{arthur.a.jaries@jyu.fi}
\affiliation{University of Jyvaskyla, Department of Physics, Accelerator laboratory, P.O. Box 35(YFL) FI-40014 University of Jyvaskyla, Finland}
\affiliation{Helsinki Institute of Physics, FI-00014, Helsinki, Finland}
\author{M.~Stryjczyk}
\email{marek.m.stryjczyk@jyu.fi}
\affiliation{University of Jyvaskyla, Department of Physics, Accelerator laboratory, P.O. Box 35(YFL) FI-40014 University of Jyvaskyla, Finland}
\author{A.~Kankainen}
\email{anu.kankainen@jyu.fi}
\affiliation{University of Jyvaskyla, Department of Physics, Accelerator laboratory, P.O. Box 35(YFL) FI-40014 University of Jyvaskyla, Finland}
\author{T.~Eronen}
\affiliation{University of Jyvaskyla, Department of Physics, Accelerator laboratory, P.O. Box 35(YFL) FI-40014 University of Jyvaskyla, Finland}
\author{O.~Beliuskina}
\affiliation{University of Jyvaskyla, Department of Physics, Accelerator laboratory, P.O. Box 35(YFL) FI-40014 University of Jyvaskyla, Finland}
\author{T.~Dickel}
\affiliation{GSI Helmholtzzentrum f\"ur Schwerionenforschung GmbH, 64291 Darmstadt, Germany}
\affiliation{II. Physikalisches Institut, Justus Liebig Universit\"at Gie{\ss}en, 35392 Gie{\ss}en, Germany}
\author{M.~Flayol}
\affiliation{Universit\'e de Bordeaux, CNRS/IN2P3, LP2I Bordeaux, UMR 5797, F-33170 Gradignan, France}
\author{Z.~Ge}
\affiliation{University of Jyvaskyla, Department of Physics, Accelerator laboratory, P.O. Box 35(YFL) FI-40014 University of Jyvaskyla, Finland}
\affiliation{GSI Helmholtzzentrum f\"ur Schwerionenforschung GmbH, 64291 Darmstadt, Germany}
\author{M.~Hukkanen}
\affiliation{University of Jyvaskyla, Department of Physics, Accelerator laboratory, P.O. Box 35(YFL) FI-40014 University of Jyvaskyla, Finland}
\affiliation{Universit\'e de Bordeaux, CNRS/IN2P3, LP2I Bordeaux, UMR 5797, F-33170 Gradignan, France}
\author{M.~Mougeot}
\affiliation{University of Jyvaskyla, Department of Physics, Accelerator laboratory, P.O. Box 35(YFL) FI-40014 University of Jyvaskyla, Finland}
\author{S.~Nikas}
\affiliation{University of Jyvaskyla, Department of Physics, Accelerator laboratory, P.O. Box 35(YFL) FI-40014 University of Jyvaskyla, Finland}
\author{I.~Pohjalainen}
\affiliation{University of Jyvaskyla, Department of Physics, Accelerator laboratory, P.O. Box 35(YFL) FI-40014 University of Jyvaskyla, Finland}
\author{A.~Raggio}
\affiliation{University of Jyvaskyla, Department of Physics, Accelerator laboratory, P.O. Box 35(YFL) FI-40014 University of Jyvaskyla, Finland}
\author{M.~Reponen}
\affiliation{University of Jyvaskyla, Department of Physics, Accelerator laboratory, P.O. Box 35(YFL) FI-40014 University of Jyvaskyla, Finland}
\author{J.~Ruotsalainen}
\affiliation{University of Jyvaskyla, Department of Physics, Accelerator laboratory, P.O. Box 35(YFL) FI-40014 University of Jyvaskyla, Finland}
\author{V.~Virtanen}
\affiliation{University of Jyvaskyla, Department of Physics, Accelerator laboratory, P.O. Box 35(YFL) FI-40014 University of Jyvaskyla, Finland}

\begin{abstract}
We report on high-precision atomic mass measurements of $^{148\text{-}153}$La and $^{151}$Ce performed with the JYFLTRAP double Penning trap using the Phase-Imaging Ion-Cyclotron-Resonance technique. The masses of $^{152,153}$La were experimentally determined for the first time. We confirm the sharp kink in the two-neutron separation energies at the neutron number ${N=93}$ in the cerium (${Z=58}$) isotopic chain. Our precision mass measurements of the most exotic neutron-rich lanthanum (${Z=57}$) isotopes reveal a unexpected sudden increase in two-neutron separation energies from ${N=92}$ to ${N=93}$. Unlike in the cerium isotopic chain, the kink is not sharp but extends to ${N=94}$ forming a prominent bump. The gain in energy is about 0.4~MeV, making it one of the strongest changes in two-neutron separation energies over the whole chart of nuclides, away from nuclear shell closures. The results, correlated with a predicted onset of quadrupole deformation for $N\geq92$, call for further studies to elucidate the structure of neutron-rich lanthanum isotopes. 
\end{abstract}

\maketitle

Penning-trap mass spectrometry can probe one of the fundamental properties of a nucleus, its mass, with high precision \cite{Eronen2016,dilling2018penning}. Nuclear binding energies $BE$ are directly related to atomic masses, and provide information on, e.g., interactions between nucleons (neutrons and protons), nuclear shell structure and deformation (see e.g. Refs.~\cite{Lunney2003,Otsuka2020}). Even subtle changes in nuclear structure can be probed via differences in binding energies along an isotopic chain, when observing abrupt deviations from the otherwise rather linear trend of the two-neutron separation energies $S_{2n}=BE(Z,N)-BE(Z,N-2)$, where $Z$ and $N$ denote the number of protons and neutrons, respectively.  

The most prominent example observed throughout the entire chart of nuclides is a sharp drop of the $S_{2n}$ values after crossing a magic neutron number, associated with a nuclear shell closure. Outside closed neutron shells, notable changes in the $S_{2n}$ trends have only been identified in three regions, located at ${N\approx32}$, ${N\approx60}$ and ${N\approx90}$. In the first region, the high-precision mass measurements of K \cite{rosenbusch2015probing}, Ca \cite{gallant2012new,wienholtz2013masses}, Sc \cite{leistenschneider2021precision, iimura2023study}, Ti and V \cite{iimura2023study} isotopes at TITAN, ISOLTRAP and BigRIPS-SLOWRI showed a sudden decrease in the $S_{2n}$ values, by about 3~MeV, between ${N=32}$ and ${N=34}$. This behavior indicates a strong shell closure at ${N=32}$ and it might be interpreted as an emergence of a new magic number \cite{Otsuka2020}. At the same time, the results of laser-spectroscopy studies contradict this interpretation \cite{GarciaRuiz2016,Koszorus2021}. 

In the $N\approx60$ region, onset of deformation was predicted \cite{Arseniev1969,Sheline1972} and experimentally discovered \cite{Cheifetz1970} already in 1970. Since then, it has been intensively studied by $\gamma$-, conversion-electron- and laser-spectroscopy, as well as mass measurements, see e.g. Ref.~\cite{Garrett2022} and references therein. The mass measurements in the neutron-rich isotopic chains between ${Z=37}$ and ${Z=42}$ \cite{Raimbault-Hartmann2002,Hager2006,Rahaman2007,hager2007precision2,Simon2012,Klawitter2016} revealed a strong kink in the $S_{2n}$ values at $N\approx60$, deviating from a linear trend by up to 2~MeV. This change has been associated with shape coexistence \cite{Garrett2022}, 
deformed and spherical orbitals crossing each other \cite{Urban2017}, and quantum phase transition \cite{togashi2016quantum}. 

In the neutron-rich rare-earth region, the onset of strong quadrupole deformation at $N\approx90$ was discovered already in the 1950s and associated with the breaking up of the $h_{11/2}$ shell \cite{Brix1952,Mottelson1955,Kleinheinz1974}. This change in nuclear structure is observed as a kink or a broader bump in the $S_{2n}$ values for isotopic chains from praseodymium (${Z=59}$) to dysprosium (${Z=66}$), increasing the $S_{2n}$ values by up to 1.5~MeV \cite{AME2020}. The kinks have been later interpreted as being due to first-stage quantum shape phase transition from spherical to prolate deformation \cite{Iachello2001}. However, recent calculations have indicated that the structure may not be as simple as traditionally predicted but can also include triaxiality, shape coexistence and octupolar deformation (see e.g. Ref.~\cite{Otsuka2019}).

The masses of neutron-rich isotopes ranging from the barium ($Z=56$) to holmium ($Z=67$) isotopic chains have been extensively studied with the JYFLTRAP Penning trap \cite{Vilen2018,Jaries2024} and the Canadian Penning trap (CPT) \cite{savard2006studies,VanSchelt2012,orford2019phase, orford2022searching}. In the case of the neutron-rich barium isotopes, a relatively flat trend in the $S_{2n}$ values is observed. While the lanthanum (${Z=57}$) and cerium (${Z=58}$) isotopic chains exhibit a small kink in the $S_{2n}$ values at $N=93$, the lack of precise experimental data hinders drawing clear conclusions on possible structural effects. The masses of $^{149,151}$La are known with a limited precision of 200 and 440~keV \cite{syntfeld2002study,knobel2016new,AME2020}, respectively, and the mass of $^{150}$La is known from a high-precision mass measurement performed at CPT \cite{orford2019phase}. In the case of $^{151}$Ce$_{93}$ isotope, the deviation from a smooth trend could be explained by the presence of a long-lived isomer proposed in the ENSDF evaluation \cite{Singh2009} but not observed in Ref.~\cite{savard2006studies}. 

In addition to poorly known mass values, the understanding of possible structural changes in the lanthanum isotopic chain around ${N\approx93}$ is hindered by the scarcity of spectroscopic information \cite{ENSDF}. The most neutron-rich lanthanum isotope studied via $\gamma$-ray spectroscopy to date remains $^{149}$La (${N=92}$) where a strong quadrupole deformation is predicted to develop  \cite{syntfeld2004first,urban2007near}. The $\gamma$-ray spectroscopy on $^{145}$La$_{88}$ and $^{147}$La$_{90}$ have shown evidence of octupole correlations \cite{urban1996octupole,zhu1999octupole,wisniewski2017parity,cardona2021ba}, similarly to the neighboring even-$Z$ chains of barium \cite{phillips1986octupole,hamilton1995new,jones1996parities,zhu1997high,hamilton1997new,sheng1997reflection,urban1997octupole,zhu1995octupole,phillips1988octupole} and cerium \cite{zhu1995octupole,phillips1988octupole,chen2006search,zhu2012observation,li2012identification}, where strong indications of octupole deformation and correlations have been observed and predicted for nuclei around $^{146}$Ba by density functional theory (DFT) calculations  \cite{Cao2020}. 

In this Letter, we report on the precise mass measurements of neutron-rich $^{148-153}$La and $^{151}$Ce isotopes performed with the JYFLTRAP double Penning-trap mass spectrometer \cite{kolhinen2004jyfltrap,eronen2014jyfltrap} using the Phase-Imaging Ion-Cyclotron-Resonance (PI-ICR) method \cite{eliseev2013phase,eliseev2014phase,nesterenko2018phase,nesterenko2021study}. Our work unveils a prominent bump in the $S_{2n}$ values at ${N\approx92-94}$ for the lanthanum isotopic chain, indicating a sudden change in the nuclear structure. We also confirm the sharp kink in the $S_{2n}$ values at $N=93$ for the cerium isotopes.

The measurements were performed at the Ion Guide Isotope Separator On-Line (IGISOL) facility \cite{moore2013towards}. The neutron-rich rare-earth isotopes were produced using a $25$-MeV proton beam with up to 20~$\mu$A of intensity impinging on a $15$-mg/cm$^2$-thick $^\text{nat}$U target. The fission fragments, mostly singly charged, were stopped in a helium-filled gas cell \cite{penttila2012fission,al2015simulations} (at about 300 mbar) before being extracted out through a sextupole ion guide \cite{karvonen2008sextupole}. The ions were further accelerated to $30q$~kV and separated based on their mass-to-charge ratio $m/q$ using a $55$~degree dipole magnet. The continuous beam was cooled and bunched with the Radiofrequency Quadrupole Cooler-Buncher \cite{nieminen2001beam} and sent to JYFLTRAP. 

\begin{figure}[h!t!b]
\includegraphics[width = 1.1\columnwidth]{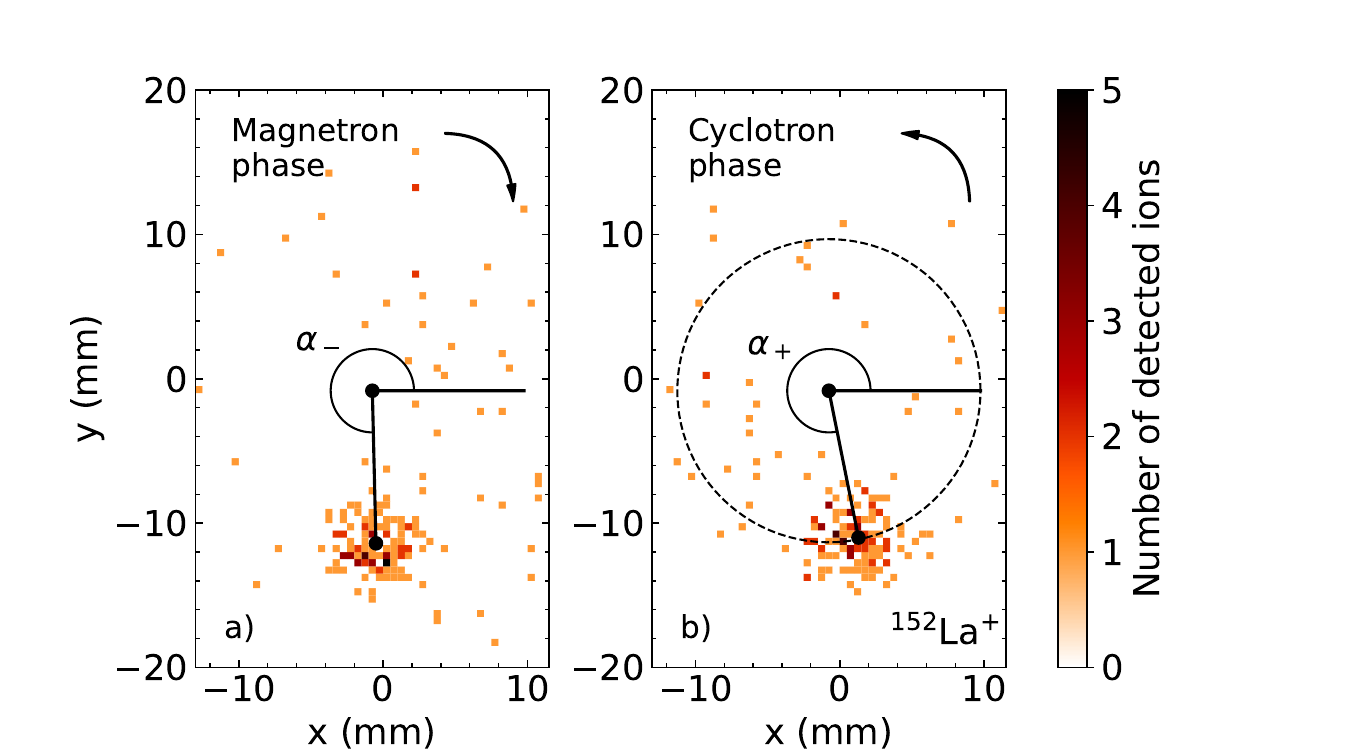}
\caption{Phase projection of \textbf{a)} the magnetron and \textbf{b)} cyclotron motion of $^{152}$La$^+$ ions on the position-sensitive detector with the PI-ICR technique using $t_\text{acc} = 92$~ms. The polar angles of the magnetron ($\alpha_-$) and cyclotron ($\alpha_+$) phase spots are indicated with an arc. The average cyclotron excitation radius is marked with the dashed circle on \textbf{b)}. \label{fig:piicr_la152} }
\end{figure}

The mass-selective buffer-gas cooling technique \cite{savard1991new} was applied in the first trap of JYFLTRAP to select the ions of interest for the mass measurements, which were performed with the PI-ICR method \cite{eliseev2013phase,eliseev2014phase,nesterenko2018phase,nesterenko2021study} in the second trap. The ion's free space cyclotron frequency $\nu_c=qB/(2\pi m)$, where $B$ is the magnetic field strength in the trap, was determined based on the detected radial eigenmotion phase spots on a position-sensitive detector after a chosen accumulation time $t_\text{acc}$ and $n_c$ revolutions \cite{eliseev2013phase,eliseev2014phase,nesterenko2018phase,nesterenko2021study}, following $\nu_c=[(\alpha_+-\alpha_-)+2\pi n_c]/(2\pi t_{acc})$, where $\alpha_{\pm}$ are the polar angles of the cyclotron ($+$) and magnetron ($-$) phase images with respect to the trap center on the detector. A typical PI-ICR figure obtained in this work is shown on Fig.~\ref{fig:piicr_la152}. 

The magnetic field strength $B$ was determined using $^{133}$Cs as a reference. The atomic mass was determined from the frequency ratio between the reference ions and ions of interest (both singly charged),   $r=\nu_{c\text{,ref}}/\nu_{c}$, as $m=r(m_\text{ref}-m_e)+m_e$, where $m_e$ is the free-electron mass. The number of detected ions per bunch was limited to up to one to reduce the effects coming from ion-ion interactions \cite{kellerbauer2003direct}. Systematic effects due to the temporal fluctuations of the magnetic field, mass dependency, and the distortion of the ions' motion projection, were taken into account, together with a residual uncertainty. The contribution of these effects is detailed in Ref.~\cite{nesterenko2021study}. A thorough description of the experimental methods used in this work can be found in Ref.~\cite{Jaries2024}. 

\begin{table*} [h!t!b]
\caption{\label{tab:table3}Summary of the nuclides studied during the mass measurement campaign using $^{133}$Cs as a reference, together with their respective half-lives $T_{1/2}$ and assigned spin-parity $J^\pi$ from NUBASE20 \cite{NUBASE2020}. The different accumulation $t_{acc}$ times with which the PI-ICR method was performed and the frequency ratios $r=\nu_\text{c,ref}/\nu_\text{c,ion}$ obtained from the measurements are reported. The extracted mass excesses from this work ME$_\text{JYFLTRAP}$ are compared to the AME20 literature values ME$_\text{AME20}$ \cite{AME2020} and the calculated differences (Diff.) are respectively shown. Values extrapolated from systematics are marked with a $\#$ symbol.}
\begin{ruledtabular}
\begin{tabular}{lllllllll}
  Nuclide & $T_{1/2}$  & $J^{\pi}$ & $t_{acc}$~(ms) & $r=\nu_\text{c,ref}/\nu_\text{c,ion}$ & $\text{ME}_\text{JYFLTRAP}$~(\text{keV}) & $\text{ME}_\text{AME20}$~(\text{keV})& \text{Diff.}~(\text{keV})\\ \hline 
\: $^{148}\text{La}$  & $1.414(25)$~s     &  (2$^-$)& 551 & \num{1.113067600(22)} & \num{-62698.5(28)} & \num{-62709(19)} & 11(19)\\
\: $^{149}\text{La}$  & $1.071(22)$~s     &  (3/2$^-$)  & 344 & \num{1.120613668(23)} & \num{-59988.3(28)} & \num{-60220(200)} & 232(200)\\
\: $^{150}\text{La}$  & $504(15)$~ms     &  (3$^+$) & 313 & \num{1.128167562(28)} & \num{-56309.4(35)} & \num{-56311.1(25)} & 2(4)\\
\: $^{151}\text{La}$  & $465(24)$~ms     &  1/2$^+\#$  & 210 & \num{1.135714229(23)} & \num{-53524.5(29)} & \num{-53310(440)}\footnotemark[1] & $-215(440)$\\
\: $^{152}\text{La}$  & $287(16)$~ms     &  2$^-\#$  & 92 & \num{1.143273524(85)} & \num{-49178(11)} & \num{-49290(300)}$\#$ & 112(300)$\#$\\
\: $^{153}\text{La}$  & $245(18)$~ms     &  1/2$^+\#$ & 83 & \num{1.15082306(30)} & \num{-46038(37)} &  &\\
\:  &      &  & 177 & \num{1.15082312(15)} & \num{-46031(18)} &  & \\
\:   &    &   &  & Weighted average: & \num{-46032(16)} & \num{-46060(300)}$\#$ & 28(300)$\#$\\
\: $^{151}\text{Ce}$  & $1.76(6)$~s     &  (3/2$^-$)  & 335 & \num{1.135652075(22)} & \num{-61219.7(27)} & \num{-61225(18)}\footnotemark[2] & 5(18)\\
\end{tabular}
\end{ruledtabular}
\footnotetext[1]{Also \num{-53542(17)} keV in Ref. \cite{kimura2024comprehensive}.}
\footnotetext[2]{Also \num{-61230(15)} keV in Ref. \cite{kimura2024comprehensive}.}
\end{table*}

The determined frequency ratios $r$ and the corresponding mass-excess (ME) values are reported in Table~\ref{tab:table3}, together with the comparison to the Atomic Mass Evaluation 2020 (AME20) \cite{AME2020}. The masses of $^{152,153}$La were experimentally determined for the first time and are in agreement with the AME20 extrapolations based on systematics. With the PI-ICR technique, the precision for all studied nuclides was greatly improved. The results are in agreement with the previously known mass values \cite{knobel2016new,savard2006studies,orford2019phase} except for $^{149}$La, which deviates by more than $1\sigma$ from the value based on a $\beta$-decay study \cite{syntfeld2002study}. We observed only one state in $^{151}$Ce, similarly to the CPT measurement \cite{savard2006studies}, thus not supporting the existence of a long-lived isomer proposed in the ENSDF evaluation~\cite{Singh2009}.

\begin{figure}[h!t!b]
\includegraphics[width =0.95\columnwidth]{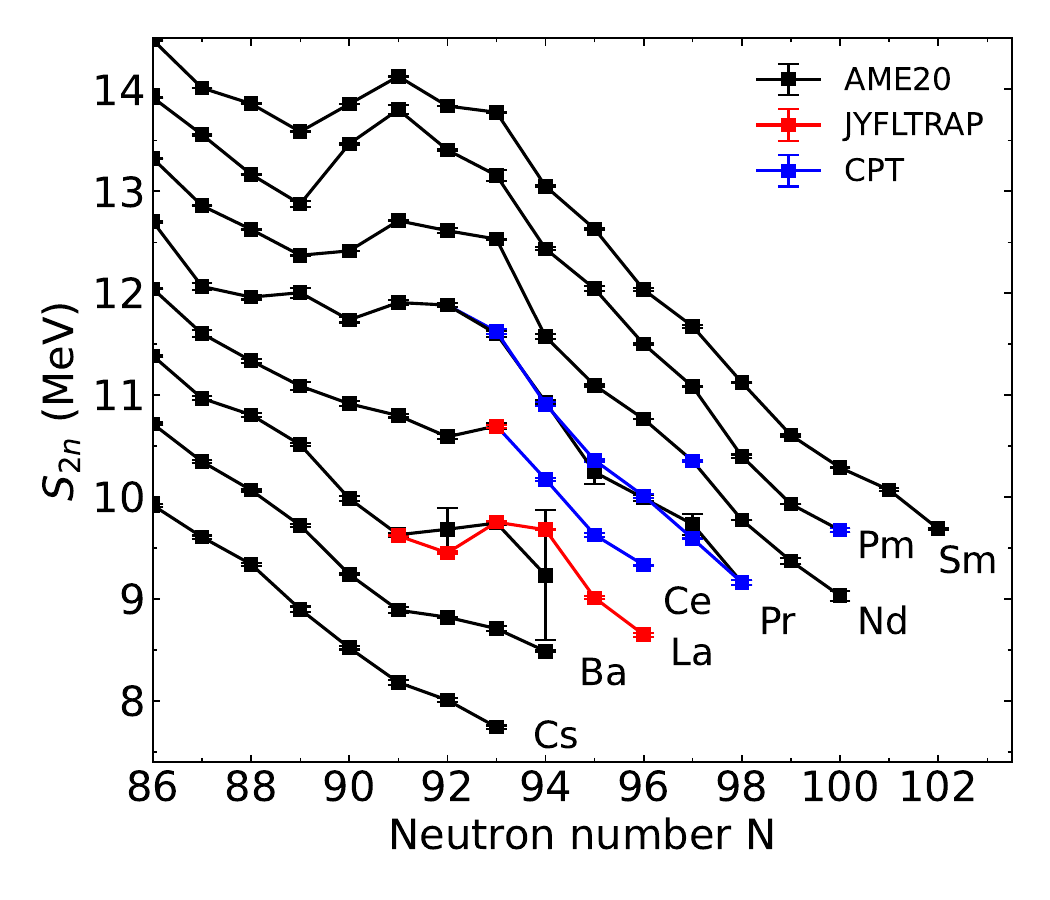}
\caption{Two-neutron separation energies $S_{2n}$ based on the new JYFLTRAP mass measurements from this work (in red) in comparison with the experimental AME20 values \cite{AME2020} (in black). For the cerium (Ce), praseodymium (Pr) and promethium (Pm) isotopic chains, also the recent CPT values are shown \cite{orford2022searching} (in blue). \label{fig:s2n}}
\end{figure}

\begin{figure}[h!t!b]
\includegraphics[width =0.95\columnwidth]{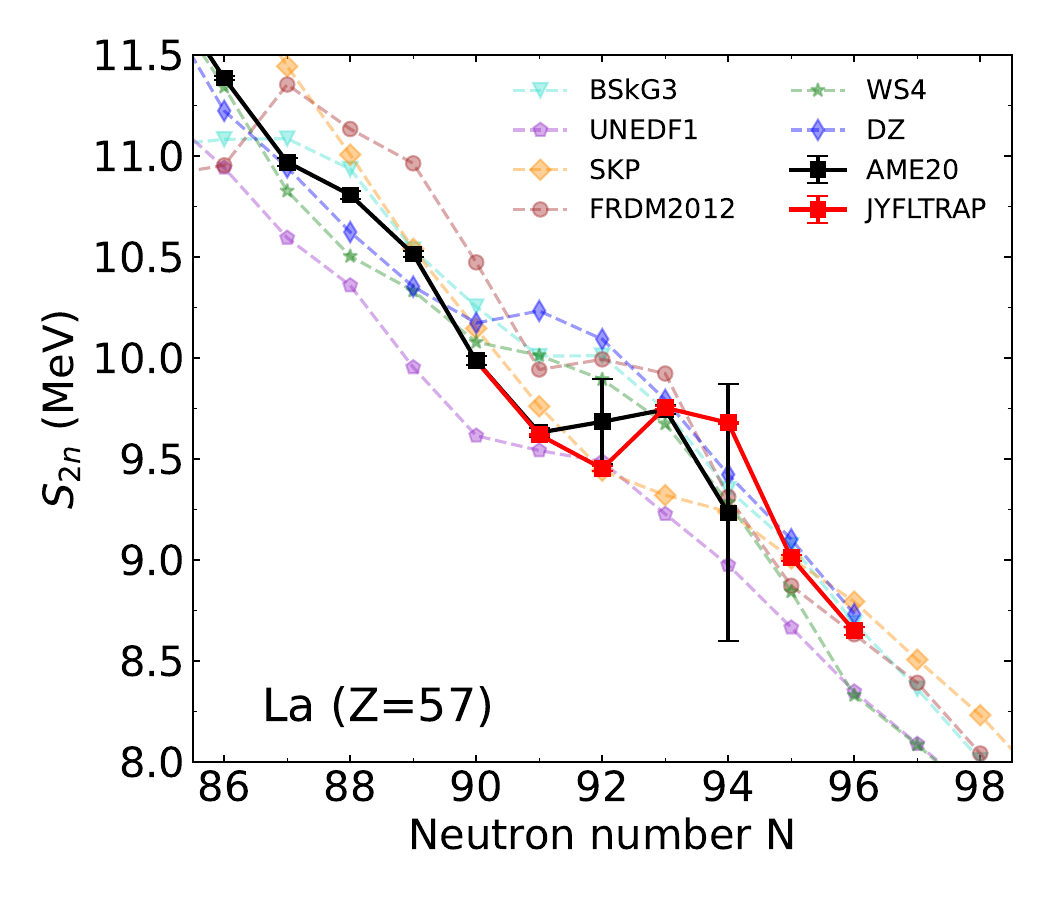}
\caption{Two-neutron separation energies $S_{2n}$ determined from this work (in red), together with the $S_{2n}$ calculated from the AME20 experimental mass values \cite{AME2020} (in black) and various mass models, namely DZ  \cite{niu2009influence}, WS4 \cite{zhao2019r}, FRDM2012 \cite{moller2016nuclear}, UNEDF1 \cite{kortelainen2012nuclear}, SKP \cite{dobaczewski1984hartree} and BSkG3 \cite{grams2023skyrme}. \label{fig:s2n_mm_comp}}
\end{figure}

\begin{figure*}
\includegraphics[width =\textwidth]{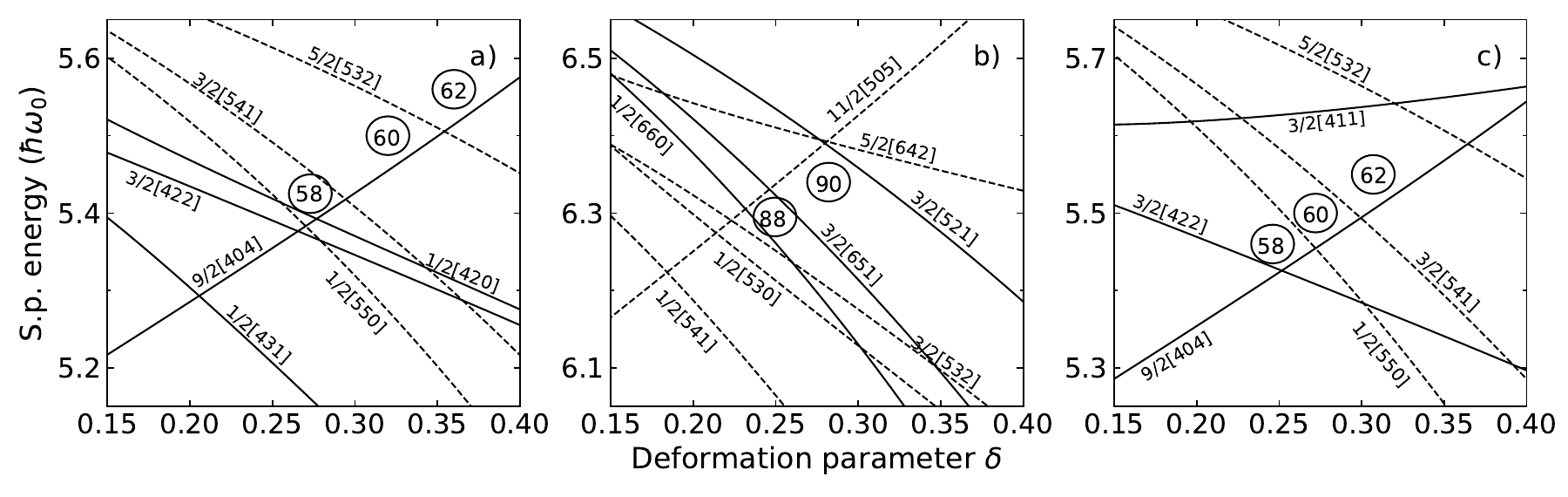}
\caption{Single-particle diagrams adapted from Ref. \cite{andersson1976nuclear} for \textbf{a)} protons around $Z=58$, \textbf{b)} neutrons around $N=88$, and \textbf{c)} neutrons around $N=60$. The levels are plotted with the deformation parameter $\delta(\approx0.95\beta)$ as defined in \cite{nilsson1955binding}. Even and odd parity are shown with solid and dashed lines, respectively. In all three regions, the upward-sloping extruder orbital crosses the downward-sloping deformation-driving orbitals. For the $N=88$ and $N=60$ regions, this has served an explanation for the observed shape coexistence and onset of ground-state deformation (see e.g. Ref.~\cite{Garrett2022,togashi2016quantum,Otsuka2019,Alexa2022}.\label{fig:nilsson}}
\end{figure*}

Figure~\ref{fig:s2n} shows two-neutron separation energies for the isotopic chains from caesium (${Z=55}$) to promethium (${Z=61}$) based on AME20 \cite{AME2020}, including our new results on $^{148-153}$La and $^{151}$Ce, as well as the recent CPT measurements for cerium, praseodymium and promethium isotopic chains \cite{orford2022searching}. The values were obtained from the mass-excesses as $S_{2n}(Z,N)=\text{ME}(Z,N-2)-\text{ME}(Z,N)+2\text{ME}_n$, where $\text{ME}_n=8071.3181(4)$~keV is the neutron mass excess \cite{AME2020}. 
While the heavier isotopic chains in the rare-earth region exhibit prominent kinks at $N\approx90$ due to onset of strong quadrupole deformation, the effect seems to vanish when moving to lower proton numbers according to AME20 values. This is in accordance with the energies of the first $2^+$ states, which decrease steeply at $N\geq88$ for the even-even nuclei in the region but much less so for the lower-$Z$ isotopic chains of cerium and barium, which reach the minimum at larger $N$ \cite{ENSDF}. 
 
In this work, we provide new, precise $S_{2n}$ values for neutron-rich lanthanum isotopes and extend the known values up to $^{153}$La (see Fig.~\ref{fig:s2n}). These reveal a pronounced $\approx$0.4~MeV bump in  the $S_{2n}$ values at neutron numbers $N=93-94$. Moreover, we can confirm the sharp small kink in the cerium $S_{2n}$ values at $N=93$ with the measurement of $^{151}$Ce.  
 
The experimental $S_{2n}$ values of lanthanum isotopes are compared to various mass models in Fig.~\ref{fig:s2n_mm_comp}. None of the models, which include both phenomenological macroscopic-microscopic and self-consistent mean-field approaches \cite{bender2003self} based on energy density functionals (EDFs), can reproduce the bump observed at $N=93-94$ but remain compatible with a rather smooth linear decrease in the $S_{2n}$ values. It should be noted that the mass models have typical root-mean-square uncertainties of a few hundred keVs and have challenges to reproduce the observed kinks in the ${N\approx60}$ and ${N\approx90}$ regions (see e.g. Refs.~\cite{Vilen2018,Hukkanen2024}). Moreover, the odd-$A$ and odd-odd nuclei are not explicitly calculated in EDF-based mass models but estimated from the pairing gap of neighboring even-even nucleus, or if calculated at the Hartree-Fock-Bogoliubov level using the quasiparticle blocking procedure, various polarization effects may appear and affect the predictions \cite{bender2003self}. Thus, the situation presented in Fig.~\ref{fig:s2n_mm_comp} is not unusual.

The $N\approx60$ and $N\approx90$ regions have been extensively studied and the prominent kinks in the $S_{2n}$ values observed in multiple isotopic chains (see e.g. Refs.~\cite{Garrett2022,Heyde2011} and references therein). At the same time, in the case of the lanthanum isotopes, our mass measurements provide the first experimental data indicating structural changes beyond ${N=92}$ and suggest that the bump at ${N=93-94}$ in the $S_{2n}$ values is a unique feature of this one element, accompanied with a smaller sharp kink for the cerium isotopes at $N=93$. While definite conclusions on the nature of the structural change cannot be drawn based on masses, spectroscopic studies can shed light on it. 

The region of neutron-rich barium to cerium isotopes is complex in structure, including both octupole and quadrupole deformations and shape coexistence. An island of octupole deformation centered around $^{146}$Ba \cite{Nazarewicz1984,Nazarewicz1992} expands to neighbouring lanthanum and cerium chains. For the latter, the spectroscopic studies suggest a gradual transition from octupolar correlations in $^{150,152}$Ce \cite{Li2012,Zhu2012} to a rigid rotor at $^{154}$Ce \cite{Urban2020}. The recent study of neutron-rich barium isotopes from the ISOLDE Decay Station indicates an increasing collectivity in $^{148-150}$Ba towards prolate deformation \cite{Lica2018}. The findings of these shape changes are further supported by theoretical calculations for the barium \cite{Lica2018} and cerium \cite{Alexa2022} chains, both indicating octupolar deformation for lighter isotopes and shape coexistence, evolving toward prolate ground-state deformation at heavier isotopes.

For the lanthanum isotopes, the spectroscopic studies of the most exotic isotope studied so far, $^{149}$La \cite{syntfeld2004first,urban2007near}, propose a transition from octupolar to prolate deformation at ${N\geq92}$. This coincides with the onset of the bump in the two-neutron separation energies, see Fig.~\ref{fig:s2n}. Urban et al. \cite{Urban2020} have discussed the potential role of the ${9/2^+[404]}$ proton extruder orbital in creating deformation in the neutron-rich ${A\approx150}$ region. The ${9/2^+[404]}$ proton extruder brings in two additional protons to the Fermi level that can be further transferred to the $3/2^-[541]$ deformation-driving down-sloping orbital (see Fig.~\ref{fig:nilsson}a) \cite{Urban2019}. The mechanism would be analogous to the neutron extruder orbitals ${11/2^-[505]}$ and ${9/2^+[404]}$ (see Figs.~\ref{fig:nilsson}b and c), which bring neutrons to the Fermi surface that can be transferred to deformation-driving intruder orbitals. 

While the neutron extruder orbitals and their role in the onset of deformation observed at ${N=88-90}$ \cite{Mottelson1955} and ${N\approx60}$ \cite{Urban2019} has been highlighted in the past, the role of the ${9/2^+[404]}$ proton extruder orbital has remained inconclusive. The prominent bump in the two-neutron separation energies of the lanthanum isotopes at $N=93-94$ and the smaller, sharp kink for the cerium isotopes at $N=93$ serve potential evidence for such a phenomenon. The location of the bump at a higher neutron number than in the heavier isotopic chains might also be explained by the proton extruder orbital. Namely, proton-neutron interactions play a crucial role in creating deformation which correlates with the number of valence protons and neutrons \cite{Casten1981,Dobaczewski1988}. The proton extruder brings in two more valence protons, requiring two more neutrons to reach the same number of valence protons and neutrons as compared to the heavier isotopic chains \cite{Urban2020}. 

In this Letter we have reported on the high-precision mass measurements of $^{148-153}$La and $^{151}$Ce performed using the PI-ICR technique with the JYFLTRAP double Penning trap. With the new mass data, we could pin down the trends in the two-neutron separation energies $S_{2n}$. Our value for $^{151}$Ce confirms the small peak at ${N=93}$ in the cerium chain, coinciding with the transitional region from octupolar to quadrupolar deformation previously studied via $\gamma$-ray spectroscopy. Our measurements extend the limits of the known lanthanum masses and reveal around 0.4-MeV bump in the $S_{2n}$ values at neutron numbers ${N=92-94}$, a unique feature within the neighboring isotopic chains not reproduced by any studied mass models. The observed change in the $S_{2n}$ trend agrees with the predicted onset of quadrupole deformation at ${N\geq92}$ that could be potentially caused by the ${9/2^+[404]}$ proton extruder orbital. This could be a plausible explanation for the localisation of the bump, as analogously neutron extruder orbitals play the main role for the other well-known kinks in two-neutron separation energies. Finally, we note that the mass evolution of neutron-rich isotopes further away from stability is crucial for astrophysical calculations, as being one of the key inputs in the determination of neutron-capture reaction rates. Specifically, the isotopes reported in this work and the possible structural changes in their vicinity were predicted to be of pivotal importance in the $r$-process \cite{Cowan2021,Surman1997,Mumpower2012,Hao2023}. A better understanding of the nuclear structure in this region is thus of broader interest for the physics community and calls for more extensive experimental and theoretical studies.

\begin{acknowledgments}
We acknowledge the funding from the European Union’s Horizon 2020 research and innovation programme under Grant Agreement No. 771036 (ERC CoG MAIDEN), No. 101057511 (EURO-LABS) and No. 861198–LISA–H2020-MSCA-ITN-2019 and from the Research Council of Finland projects No. 354589, No. 295207, No. 306980, No. 327629 and No. 354968. The financial support provided by the Vilho, Yrjö and Kalle Väisälä Foundation is acknowledged by J. Ruotsalainen. We thank Markus Kortelainen for insightful discussions.
\end{acknowledgments}

\bibliography{bibliography}

\end{document}